\documentclass[12pt]{article}
\usepackage{epsf,amsfonts,amssymb,epsfig,amsmath}
\renewcommand{\text}[1]{#1}

\newcommand{\be}{\begin{equation}}
\newcommand{\ee}{\end{equation}}
\newcommand{\ben}{\begin{displaymath}}
\newcommand{\een}{\end{displaymath}}
\newcommand{\bea}{\begin{eqnarray}}
\newcommand{\eea}{\end{eqnarray}}
\newcommand{\bean}{\begin{eqnarray*}}
\newcommand{\eean}{\end{eqnarray*}}
\newcommand{\nn}{\nonumber \\}
\newcommand{\ba}{\begin{array}}
\newcommand{\ea}{\end{array}}
\newcommand{\bi}{\begin{itemize}}
\newcommand{\ei}{\end{itemize}}

\newcommand{\reef}[1]{(\ref{#1})}

\def\a{\alpha}

\def\b{\beta}
\def\c{\sigma}

\newcommand{\bbR}{{\mathbb{R}}}

\begin{document}

\makeatletter
\renewcommand{\theequation}{\thesection.\arabic{equation}}
\@addtoreset{equation}{section}
\makeatother

\baselineskip 18pt

\begin{titlepage}

\vfill

\begin{flushright}
Imperial/TP/2007/JG/03\\
\end{flushright}

\vfill

\begin{center}
   \baselineskip=16pt
  {\Large\bf Geometries with Killing Spinors and Supersymmetric $AdS$ Solutions}
   \vskip 2cm
      Jerome P. Gauntlett$^1$ and Nakwoo Kim$^2$\\
   \vskip .6cm
      \begin{small}
      $^1$\textit{Theoretical Physics Group, Blackett Laboratory, \\
        Imperial College, London SW7 2AZ, U.K.}
        \end{small}\\*[.6cm]
      \begin{small}
      $^1$\textit{The Institute for Mathematical Sciences, \\
        Imperial College, London SW7 2PE, U.K.}
        \end{small}\\*[.6cm]
      \begin{small}
     $^2$\textit{Department of Physics and Research Institute of
        Basic Science, \\
        Kyung Hee University, Seoul 130-701, Korea}
        \end{small}
   \end{center}

\vfill

\begin{center}
\textbf{Abstract}
\end{center}

\begin{quote}
The seven and nine dimensional geometries associated with 
certain classes of supersymmetric $AdS_3$ and $AdS_2$ solutions of 
type IIB and $D=11$ supergravity, respectively, have many similarities
with Sasaki-Einstein geometry. We further elucidate their properties and
also generalise them to higher odd dimensions by
introducing a new class of complex geometries 
in $2n+2$ dimensions, specified by a Riemannian 
metric, a scalar field and a closed three-form, 
which admit a particular kind of Killing spinor. 
In particular, for $n\ge 3$, we show that when
the geometry in $2n+2$ dimensions is a cone 
we obtain a class of geometries in $2n+1$ dimensions, specified by a Riemannian
metric, a scalar field and a closed two-form, 
which includes the seven and nine-dimensional geometries mentioned above
when $n=3,4$, respectively. 
We also consider various ansatz for the geometries 
and construct infinite classes of explicit examples for all $n$.

\end{quote}

\vfill

\end{titlepage}
\setcounter{equation}{0}


\section{Introduction}

An interesting class of geometries in seven and nine dimensions
was recently discovered in \cite{nak1,nak2} and further explored in \cite{gkw}. 
The geometries are specified by a Riemannian metric, a scalar field $B$ and a
closed two-form $F$ and they admit Killing spinors of a certain type.
The seven dimensional geometries give rise to supersymmetric solutions of type IIB supergravity
with a three dimensional anti-de-Sitter space ($AdS_3$) factor and these
are dual to supersymmetric conformal field theories (SCFTs) with $(0,2)$ supersymmetry in
two-dimensions. Similarly, the nine-dimensional geometries give rise to supersymmetric
solutions of $D=11$ supergravity with $AdS_2$ factors and these are dual to 
superconformal quantum mechanics with two supercharges. 

This geometry in $2n+1$ dimensions, with $n=3,4$,  is strikingly
similar to Sasaki-Einstein geometry. 
In particular, they both have a Killing Reeb vector of constant norm and define a $U(n)$ or 
metric contact structure. The Killing vector defines a natural foliation 
and in the Sasaki-Einstein case
the metric transverse to these orbits is K\"ahler and Einstein, while for the 
geometry considered in \cite{nak1,nak2} it is K\"ahler and in addition satisfies
\be\label{inakeq}
\Box R -\tfrac{1}{2} R^2 + R_{ij}R^{ij} = 0,
\ee
where $R$ and $R_{ij}$ are the Ricci-scalar and Ricci-tensor, 
respectively, of the transverse metric and we also 
demand\footnote{Note that when $R<0$ we can construct Lorentzian geometries
which for $n=3,4$ give rise to solutions of type IIB supergravity and
$D=11$ supergravity with an $S^3$ and $S^2$ factor, respectively, as
described in \cite{gkw}.} that $R>0$. Moreover, locally, the whole geometry can be reconstructed from
a local K\"ahler metric in $2n$-dimensions satisfying \reef{inakeq}.

Recall that a succinct definition of a Sasaki-Einstein metric in $2n+1$ dimensions
is that the corresponding cone metric in $2n+2$ dimensions, 
with base given by the Sasaki-Einstein metric, is Ricci-flat 
and K\"ahler, or equivalently has $SU(n+1)$ holonomy. 
A metric with $SU(n+1)$ holonomy has an $SU(n+1)$ structure, specified by a 
fundamental two-form
$J$ and an $(n+1,0)$-form $\Omega$ (which together define a metric), 
with vanishing intrinsic torsion,
\be
d J=d\Omega=0.
\ee
Equivalently it can be characterised as admitting covariantly constant spinors.

In this paper we will determine the analogous statements for
the geometries studied in \cite{nak1,nak2} and furthermore generalise 
the geometry from $n=3,4$ to all $n\ge 3$. We find that the analogue of a metric
with $SU(n+1)$ holonomy is a geometry specified by a metric, a scalar field $\phi$ 
and a closed three-form $f$
with an $SU(n+1)$ structure $(J,\Omega)$ satisfying 
\bea\label{icon}
d[e^{n\phi}\Omega]&=&0\nn
d[e^{2(n-1)\phi}J^n]&=&0\nn
d[e^{2\phi}J]&=&f
\eea
and in addition
\be\label{ipink}
d\left[e^{2(n-3)\phi}*_{2n+2}f\right]=0,
\ee
where $*_{2n+2}$ is the Hodge dual using the $(2n+2)$-dimensional metric
defined by the $SU(n+1)$ structure.
Note that \reef{icon} imply that the almost complex structure associated
with the $SU(n+1)$ structure is integrable.
We will show that complex geometries satisfying \reef{icon} are equivalent
to geometries that admit a certain type of Killing spinor.
Furthermore, by analysing the integrability conditions for the 
Killing spinor equations, and in addition imposing \reef{ipink},
we will also determine the equations of motion satisfied by the metric, the
scalar and the three-form, which are the analogue of the property of
Ricci-flatness in the case of $SU(n+1)$ holonomy.

If we demand that the geometry in $2n+2$ dimensions satisfying
\reef{icon}, \reef{ipink} is a metric cone and 
with the scalar and the three-form having a specific scaling:
\bea\label{iscal}
ds^2_{2n+2}&=&dr^2+r^2 ds^2_{2n+1}\nn
e^{-2\phi}&=&r^{\frac{2(n-1)}{n-2}}e^B\nn
f&=&r^{\frac{n}{2-n}}dr\wedge F,
\eea
we will show that we 
obtain a geometry in $2n+1$ dimensions specified by a metric, with line element $ds^2_{2n+1}$, a scalar field, $B$, and
a closed two-form, $F$, all independent of the co-ordinate $r$,
which for $n=3,4$ is precisely equivalent to the geometry of \cite{nak1,nak2}. 
We show that for all $n$ the $2n+1$ dimensional metric
has a Killing vector of constant norm and that the transverse metric
is K\"ahler and satisfies \reef{inakeq}. For all $n$,
locally, the whole geometry can be reconstructed from
a local K\"ahler metric in $2n$-dimensions satisfying \reef{inakeq}.
We determine the kind of Killing spinors that the geometry in $2n+1$ dimensions admit 
and also the kind of equations of motion that are satisfied by the metric, the scalar field $B$ and
the two-form $F$, which are the analogue of
the Einstein condition in the case of Sasaki-Einstein geometry.

Note that in $2n+2$ dimensions we are generalising the notion of $SU(n+1)$ holonomy
in the sense that if we set $f=\phi=0$ in \reef{icon}, \reef{ipink} we clearly
return to the case of $SU(n+1)$ holonomy. However, on the base of the cone in
$2n+1$ dimensions, defined by \reef{iscal}, we are not generalising the notion
of Sasaki-Einstein geometry: for example \reef{inakeq} is not satisfied for
Einstein metrics for $n\ge 3$.

When $n=3$ an eight dimensional geometry satisfying \reef{icon}, \reef{ipink}
gives rise to a supersymmetric solution of type IIB supergravity which is a warped product
of $\bbR^{1,1}$ with the eight-dimensional geometry.
Assuming that the eight dimensional geometry is a cone
as in \reef{iscal} we recover the type IIB $AdS_3$ solutions of \cite{nak1}.
Similarly, as discussed in \cite{Mac Conamhna:2006nb},
when $n=4$, a ten-dimensional geometry satisfying \reef{icon}, 
\reef{ipink} gives rise to a supersymmetric solution of 
$D=11$ supergravity which is a warped product of $\bbR$ with the ten-dimensional geometry. 
If the ten dimensional geometry is
a cone as in \reef{iscal} we recover the $D=11$ $AdS_2$ solutions of \cite{nak2}.

We do not know of any physical application for the geometry when $n\ge 5$.
However, it is possible that the geometries in $2n+1$ dimensions, with $n\ge 5$, 
inherit some properties dictated by physics for the seven and/or nine dimensional cases. 
This is by analogy with the Sasaki-Einstein case. 
Recall that five-dimensional Sasaki-Einstein geometries, $SE_5$, give rise
to $AdS_5\times SE_5$ solutions of type IIB supergravity. These solutions 
are dual to $N=1$ supersymmetric conformal field theories
(SCFTs) in four spacetime dimensions and such SCFTs exhibit the 
phenomenon of $a$-maximisation \cite{Intriligator:2003jj}. 
Motivated by this observation, it was proven in \cite{Martelli:2005tp,Martelli:2006yb} 
that the volume of Sasaki-Einstein manifolds in {\it any dimension} 
satisfies a variational principle.

Sections 2 and 3 of this paper will be devoted to expanding on the above discussion.
In the subsequent sections we will then consider various ansatz in order to find explicit examples.
In section 4 we will construct explicit examples of the geometries in $2n+1$ dimensions.
This is a direct analogue of the explicit construction of Sasaki-Einstein metrics that was carried out in
\cite{Gauntlett:2004yd,Gauntlett:2004hh} and generalises the analysis of 
\cite{gkw} from $n=3,4$ to all $n\ge 3$. 
More specifically, we construct explicit local K\"ahler metrics in $2n$-dimensions satisfying 
\reef{inakeq}, by considering local metrics on line bundles over positively curved K\"ahler-Einstein manifolds in $2n-2$ dimensions. 
We then argue that for each choice of K\"ahler-Einstein manifold these 
lead to countably infinite classes of smooth,
compact and simply connected globally defined geometries in $2n+1$ dimensions.

Section 5 will present an ansatz for geometries in $2n+2$ dimensions that depend on a number of
functions of one variable. We show that the ansatz includes the simple case of a cone
over a $2n+1$ dimensional geometry with the corresponding $2n$-dimensional K\"ahler manifold satisfying
\reef{inakeq} being a product of K\"ahler-Einstein spaces. Such $2n+1$-dimensional geometries, for $n=3,4$ 
were studied in \cite{gkw}.
We also show that the ansatz includes singular non-compact Calabi-Yau geometries, some of which were
discussed in \cite{Cvetic:2000db}. 
For $n=3$ the ansatz also incorporates a known solution in type IIB supergravity 
that describes an interpolation between a solution with an $AdS_5$ factor and
a solution with an $AdS_3\times H_2/\Gamma$ factor, where 
$H_2$ is the hyperbolic plane and $\Gamma$ is a discrete group of isometries
\cite{Klemm:2000nj,Maldacena:2000mw}. 
Similarly, for $n=4$ the ansatz covers a known solution in $D=11$ supergravity 
that describes an interpolation between a solution with an $AdS_4$ factor and
a solution with an $AdS_2\times H_2/\Gamma$ factor \cite{Caldarelli:1998hg,Gauntlett:2001qs}. 

In section 6 we will consider an ansatz for the geometries in $2n+1$ dimensions
which is inspired by the work of \cite{Lin:2004nb}. 
The resulting system boils down to solving a differential
equation for a function $D$ of three variables, $x^1,x^2, z$. 
For $n=3$ the equation is linear,
as in \cite{Lin:2004nb}, and can be explicitly solved. 
For $n\ge 4$ the equation is
\be
\Delta D + z^{\frac{n-4}{n-3}}\partial_z^2 e^D  = 0,
\ee
where $\Delta=\partial_1^2+\partial_2^2$.
For $n=4$ this is equivalent to the continuous Toda equation as in 
\cite{Lin:2004nb}. We don't know whether the equation is an integrable system
for $n\ge 5$. 

Section 7 briefly concludes.

\section{Geometry in $2n+2$ Dimensions}

The geometry in $2n+2$ dimensions that we will be interested in is specified by a 
Riemannian metric, $g$, a scalar field, $\phi$, and a closed three-form, $f$:
\be\label{pinkone}
df=0.
\ee
We are interested
in such geometries that admit a solution to the Killing spinor equations:
\bea\label{kspin}
\left[\gamma^\a\nabla_\a\phi   + \frac{i}{12} e^{-2\phi}f_{\c_1\c_2\c_3} \gamma^{\c_1\c_2\c_3}\right]\epsilon &=& 0
\nn
\left[\nabla_\a  -\frac{i}{24} e^{-2\phi}f_{\c_1\c_2\c_3} \gamma_{\a}{}^{\c_1\c_2\c_3}\right] \epsilon &=& 0,
\eea
where $\epsilon$ is a $Spin(2n+2)$ spinor, the gamma-matrices, $\gamma^\alpha$, generate the
Clifford algebra $Cliff(2n+2)$: $\{\gamma_\alpha,\gamma_\beta\}=2g_{\alpha\beta}$
and the indices $\alpha,\sigma,\dots$ run from $1$ to $2n+2$.
We will be particularly interested in geometries admitting such Killing spinors that in
addition satisfy the following equation of motion for $f$:
\be\label{pink}
d\left[e^{2(n-3)\phi}*_{2n+2}f\right]=0.
\ee

We now argue that any geometry satisfying \reef{pinkone}, \reef{kspin}, \reef{pink}
is also a solution to the following equations:
\bea\label{eom}
E_{\a\b}&=&0\nn
\nabla^2\phi+2(n-1)(\nabla \phi)^2-\frac{1}{2}e^{-4\phi}f^2&=&0,
\eea
where we have defined 
\be
E_{\a\b}\equiv R_{\a\b}-2(n-1)\nabla_{\a\b}\phi+2(n-2)\nabla_\a \phi\nabla_\b \phi
+\frac{1}{4}e^{-4\phi}f_{\a\c_1\c_2}f_\b{}^{\c_1\c_2}-\frac{1}{2}g_{\a\b}e^{-4\phi}f^2.
\ee
To see this we follow an argument of \cite{Gauntlett:2002fz}.
Specifically, the integrability conditions for the 
Killing spinor equations can be used to show that
\bea\label{one}
E_{\b\c}\gamma^\c\epsilon
=-\frac{i}{48}e^{-2\phi}df_{\c_1\c_2\c_3\c_4}\gamma_{\b}{}^{\c_1\c_2\c_3\c_4}\epsilon
-\frac{i}{4}e^{2(2-n)\phi}\nabla_\a(e^{-2(3-n)\phi}f^\a{}_{\c_1\c_2})\gamma_\b{}^{\c_1\c_2}\epsilon
\eea
and
\bea\label{two}
\left[\nabla^2\phi+2(n-1)(\nabla\phi)^2-\frac{1}{2}e^{-4\phi}f^2\right]\epsilon=\nn
-\frac{i}{48}e^{-2\phi}df_{\c_1\c_2\c_3\c_4}\gamma^{\c_1\c_2\c_3\c_4}\epsilon
-\frac{1}{4}ie^{2(2-n)\phi}\nabla_\a(e^{-2(3-n)\phi}f^\a{}_{\c_1\c_2})\gamma^{\c_1\c_2}\epsilon.
\eea
If we now impose \reef{pinkone}, \reef{pink}, we immediately deduce from \reef{two} 
that the scalar equation of motion in \reef{eom} is satisfied. From \reef{one} we similarly deduce that 
$E_{\a\b}\gamma^\b\epsilon=0$, but on a Riemannian manifold this implies that $E_{\a\b}=0$.

Observe that \reef{pink}, \reef{eom} are equations of motion that
can be derived by varying an action with Lagrangian density given by 
\be
{\cal L}_{2n+2}=e^{2(n-1)\phi}\left[R+2n(2n-3)(\nabla \phi)^2+\frac{1}{2}e^{-4\phi}f^2\right].
\ee
Here we have defined $f^2\equiv (1/3!)f_{\a_1\a_2\a_3}f^{\a_1\a_2\a_3}$ and we are thinking
of the action as being a functional of the metric, the scalar $\phi$ and a two-form potential 
$b$ with $f=db$.

We next observe that the only compact solutions 
to the equations of motion \reef{pink}, \reef{eom}
are Ricci-flat manifolds. To see this note that
the scalar equation of motion implies that
\be
\nabla^2[e^{2(n-1)\phi}]=2(n-1)e^{2(n-3)\phi}f^2.
\ee
Integrating this over a compact manifold we deduce for $n\ge 2$ that $f=0$.
The scalar equation of motion in \reef{eom} 
then implies that $\phi=0$. A similar argument works
for $n=1$ also. In section 3 we will focus on non-compact cone geometries
which can have compact base spaces.

\subsection{$SU(n+1)$ structure}
We now restrict our considerations to solutions of the Killing spinor equations \reef{kspin}
where the Killing spinor $\epsilon$ is a Weyl spinor. More specifically, we demand that $\epsilon$ 
is no-where vanishing and has isotropy group $SU(n+1)\subset Spin(2n+2)$.
In other words we demand that the Killing spinor fixes a globally defined $SU(n+1)$-structure.

We first observe that the Killing spinor equations \reef{kspin} imply that
$\bar\epsilon\epsilon$ is a constant. We will fix the normalisation by imposing
$\bar\epsilon\epsilon=1$.  The $SU(n+1)$ structure is specified by a fundamental
two-form $J$ and an $(n+1,0)$-form $\Omega$ both of which 
can be constructed as bi-linears in $\epsilon$:
\bea
J_{\a\b}&=&-i\bar\epsilon\gamma_{\a\b}\epsilon\nn
\Omega_{\a_1\dots \a_{n+1}}&=&\bar\epsilon^c\gamma_{\a_1\cdots \a_{n+1}}\epsilon,
\eea
where $\epsilon^c$ is the spinor conjugate to $\epsilon$.
Recall that $(J,\Omega)$ define a metric and an almost complex structure.
After some detailed calculations, we find that the Killing spinor equations \reef{kspin} imply that the
$SU(n+1)$ structure must satisfy
\bea\label{con}
d[e^{n\phi}\Omega]&=&0\nn
d[e^{2(n-1)\phi}J^n]&=&0\nn
d[e^{2\phi}J]&=&f.
\eea
These equations account for all of the intrinsic torsion modules of the 
$SU(n+1)$ structure. In particular,
using the notation of \cite{Gauntlett:2003cy}, 
the first equation in \reef{con} says that the torsion modules
$W_1=W_2=0$, which implies that the manifold is complex (i.e. that the almost
complex structure is integrable), and that
the Lee form $W_5\propto d\phi$. The second equation in \reef{con}
says that the Lee form $W_4\propto d\phi$. The third equation in \reef{con}
relates $W_3$ and $W_4$ to the three-form $f$.

We have argued that \reef{con} are necessary conditions
for solutions of the Killing spinor equations \reef{kspin} with spinors that
define an $SU(n+1)$ structure. They are also sufficient. In particular given an $SU(n+1)$ structure
satisfying the first two conditions in \reef{con}, one can
extract $d\phi$ from the torsion modules $W_4$ or $W_5$ and obtain a three-form $f$ via 
the last equation. Following the same type of argument as that discussed after equation (4.23) of
\cite{Gauntlett:2002fz} we conclude that there will be an $SU(n+1)$ invariant 
Weyl spinor that solves the Killing spinor equations \reef{kspin}.

Clearly the Bianchi identity for $f$, \reef{pinkone}, is automatically 
implied by \reef{con}. Thus in light of the integrability argument made in the previous 
subsection, if we also impose the equation of motion for $f$, \reef{pink},
then we deduce that all of the equations of motion \reef{eom} are satisfied.
Also observe that we are describing a generalisation of manifolds with special holonomy
$SU(n+1)$. In particular, if $\phi=f=0$, we are demanding
the existence of $SU(n+1)$ invariant covariantly constant spinors, in other words
geometries with $SU(n+1)$ holonomy, and \reef{con} reduces to the usual conditions $dJ=d\Omega=0$.

The geometries with Killing spinors that we are describing generalise 
a certain class of supersymmetric 
solutions of type IIB and $D=11$ supergravity. Specifically, we have 
checked\footnote{It was recently shown in \cite{Gauntlett:2007ph} that this result
can also be obtained by considering a restricted class of solutions
analysed in \cite{Gran:2007ps}.}
that the geometry with $n=3$ satisfying \reef{con} and \reef{pink}
gives rise to a supersymmetric solution of type IIB supergravity of the form:
\bea\label{10ant}
ds^2&=&e^{\phi}[ds^2(\bbR^{1,1})+ds^2_{8}]\nn
F_5&=&-\frac{1}{4}[Vol(\bbR^{1,1})\wedge f- *_8 f],
\eea
where $F_5$ is the self-dual five form. These solutions preserve 
$(0,2)$ supersymmetry with respect to $\bbR^{1,1}$.
Similarly, the $n=4$ geometry satisfying \reef{con} and \reef{pink}
gives \cite{Mac Conamhna:2006nb} the following supersymmetric solution of
$D=11$ supergravity
\bea\label{11ant}
ds^2&=&e^{4\phi/3}[-dt^2+ds^2_{10}]\nn
G_4&=&dt\wedge f.
\eea
These solutions preserve two supercharges.
For both of these cases flux quantisation in the supergravity theory
implies that the periods of $f$ should be rational\footnote{Actually, to be more
precise, the quantisation condition on the four-form is slightly different: see \cite{wit}.}. 
One might consider demanding that this condition holds for general $n$.

In the next section we will assume that the metric in $2n+2$
dimensions is a metric cone, as well as imposing additional assumptions on $\phi,f$, 
and study the corresponding geometry on the $2n+1$-dimensional base of the cone. 
Before doing that, let us conclude with two comments which will play no role in the sequel.
Firstly, we observe that the Killing spinor equations in \reef{kspin},
for arbitrary spinor, can be equivalently written:
\bea
\gamma^\a\nabla_\a\phi \epsilon  + \frac{i}{12} e^{-2\phi}f_{\c_1\c_2\c_3} \gamma^{\c_1\c_2\c_3}\epsilon &=& 0\nn
\left[\nabla_\a
+\frac{1}{2}\nabla_\a\phi+\frac{1}{2}\nabla_\b\phi\gamma_{\a}{}^\b+\frac{i}{8}e^{-2\phi}f_{\a\c_1\c_2}\gamma^{\c_1\c_2}\right]\epsilon&=&0.
\eea
If we now introduce the rescaled metric $\tilde g=e^{2\phi}g$ and the rescaled spinor $\tilde \epsilon=e^{\phi/2}\epsilon$
the Killing spinor equations become (dropping tildes)
\bea
\gamma^\a\nabla_\a\phi \epsilon  + \frac{i}{12} f_{\c_1\c_2\c_3}\gamma^{\c_1\c_2\c_3}\epsilon &=& 0
\nn
\nabla_\a \epsilon  +\frac{i}{8}  f_{\a\c_1\c_2}\gamma^{\c_1\c_2} \epsilon &=& 0.
\eea
Interestingly these are just the Killing spinor equations that arise in the
common NS-NS sector of supergravity (see \cite{strom,hull} and e.g. \cite{Gauntlett:2002sc,Gauntlett:2003cy}) 
with imaginary 3-form flux $H=i f$ and dilaton $\Phi=\phi$.

Although we will be focussing on $n\ge 3$ in the remainder, the second comment 
concerns the $n=2$ case. If we let $\epsilon$ be a chiral spinor: $i\gamma_7\epsilon=\epsilon$ where $\gamma_7=\gamma_{123456}$
and define $H=e^{-2\phi}*_6f$ and the dilaton $\Phi=-\phi$, then we
find the Killing spinor equations are exactly the same as in the common NS-NS sector 
in six dimensions.

\section{Geometry in $2n+1$ Dimensions}

We now restrict our considerations to $n\ge 3$ and take the $2n+2$ dimensional metric of the last section
to be a cone metric:
\be\label{cone1}
ds^2_{2n+2}=dr^2+r^2 ds^2_{2n+1},
\ee
where $ds^2_{2n+1}$ is independent of $r$.
We also demand that the scalar field $e^{2\phi}$ and the
three-form $f$ have the following dependence on $r$
\bea\label{cone2}
e^{-2\phi}&=&r^{\frac{2(n-1)}{n-2}}e^B\nn
f&=&r^{\frac{n}{2-n}}dr\wedge F.
\eea
where the scalar field $B$ and the closed two-form $F$ are independent of $r$.
We are interested in the geometry in $2n+1$ dimensions on the link
$L$, defined to be the surface $r=1$ on the cone, with 
metric whose line element is $ds^2_{2n+1}$, scalar field $B$ and closed two-form $F$.

We first observe that the equations of motion in $2n+2$ dimensions
given in \reef{pinkone}, \reef{eom} give rise to the following equations of motion in $2n+1$ 
dimensions 
\bea\label{eom2}
R_{ab}+(n-1)\nabla_{ab}B+\frac{(n-2)}{2}\nabla_a B\nabla_b B+g_{ab}\frac{2}{n-2}
+\frac{1}{2}e^{2B}F_{ac}F_b{}^c-\frac{1}{4}g_{ab}F^2=0\nn
\nabla^2B-(n-1)(\nabla B)^2-\frac{4(n-1)}{(n-2)^2}       +\frac{e^{2B}}{2}F^2=0\nn
d\left[e^{(3-n)B}*_{2n+1}F\right]=0
\eea
where $F^2\equiv F_{ab}F^{ab}$.
These equations of motion can be derived from an action with Lagrangian 
given by\footnote{Note for $n=3$ that if we change the sign of the last two terms in
this Lagrangian, we obtain a Lagrangian equivalent to one considered in \cite{Chen:2007du}.}
\be
{\cal L}_{2n+1}=e^{(1-n)B}\left[R+\frac{n(2n-3)}{2}(\nabla B)^2+\frac{1}{4}e^{2B}F^2-\frac{2n}{(n-2)^2}\right].
\ee
Here we are thinking of the action as being a functional of the metric, the scalar $B$ and a
one-form potential $A$ with $F=dA$.

The Killing spinor equations in $2n+2$ dimensions \reef{kspin} 
give rise to Killing spinor equations in $2n+1$ dimensions.
The generators $\gamma_\a$ of $Cliff(2n+2)$ can be written
\bea
\gamma_a&=&\Gamma_a\otimes\sigma_1\quad a=1,\dots,2n+1\nn
\gamma_r&=&1\otimes\sigma_2,
\eea
where $\Gamma_a$ generate $Cliff(2n+1)$ and $\sigma_1, \sigma_2$ are Pauli matrices.
For definiteness, when $n$ is odd 
we take $\Gamma_1\dots \Gamma_{2n+1}=-i$ and the chirality
operator in $2n+2$ dimensions as $\gamma_1\dots\gamma_{2n+1}\gamma_r=1\otimes\sigma_3$. When $n$ is even we
take $\Gamma_1\dots \Gamma_{2n+1}=-1$ and the chirality
operator in $2n+2$ dimensions as $i\gamma_1\dots\gamma_{2n+1}\gamma_r=1\otimes\sigma_3$. In both cases, then,
a positive chirality spinor in $2n+2$ dimensions can be written as 
$\epsilon=(\eta,0)$ where $\eta$ is a spinor in $2n+1$ dimensions. We then find
that substituting \reef{cone1} and \reef{cone2} into \reef{kspin} leads to
\bea\label{kspin2}
\left[\Gamma^a\nabla_aB + i\frac{2(n-1)}{n-2}+\frac{1}{2}e^BF_{ab}\Gamma^{ab}\right]\eta &=& 0
\nn
\left[\nabla_c  +\frac{i}{2}\Gamma_c+\frac{1}{8}\ e^{B}F_{ab} \Gamma_{c}{}^{ab}\right] \eta &=& 0.
\eea
Using a result of the last section, we also conclude that
if we have a solution to these Killing spinor equations and in addition we impose the
Bianchi identity, $dF=0$, and the equation of motion for the two-form
\be\label{Feom2}
d\left[e^{(3-n)B}*_{2n+1}F\right]=0,
\ee 
then all of the equations of motion in \reef{eom2} will be satisfied.

For the $n=3$ case a solution to the Killing spinor equations \reef{kspin2} and \reef{Feom2}
give rise to a supersymmetric type IIB solution with an $AdS_3$ factor of the form 
\bea
ds^2&=&e^{-B/2}[ds^2(AdS_3)+ds^2_{7}]\nn
F_5&=&-\frac{1}{4}[Vol(AdS_3)\wedge F- *_7 F],
\eea
while for the $n=4$ case we can obtain the following solution of
$D=11$ supergravity with an $AdS_2$ factor
\bea
ds^2&=&e^{-2B/3}[ds^2(AdS_2)+ds^2_{9}]\nn
G_4&=&Vol(AdS_2)\wedge F.
\eea
In making the comparison with \cite{nak1,nak2,gkw} one should identify
$B=\frac{2(n-1)}{2-n}A$ and in the IIB case we have $F^{here}=-4F^{there}$.

We can analyse the geometries in $2n+1$ dimensions that admit solutions to the Killing spinor
equations \reef{kspin} with $SU(n+1)$ invariant spinors
in two ways. We can either directly analyse
the Killing spinor equations \reef{kspin2}, generalising the analysis of \cite{nak1,nak2}, or 
equivalently, we can reduce the $SU(n+1)$ structure in $2n+2$ dimensions satisfying
\reef{con} that we 
discussed in the last section. Let us consider the latter approach.

We first stay on the cone. We introduce the Reeb vector field $\xi$ defined by
\be
\xi^\alpha=J^\alpha{}_\beta (r\partial_r)^\beta,
\ee
which has norm squared given by $r^2$.
In this expression $J$ refers to the integrable complex structure on the cone
obtained by raising an index on the two-form $J$; we hope that using
the same letter for both does not cause confusion. 
We also define the one form $\eta$ on the cone:
\be
\eta_\alpha=J_\alpha{}^\beta (\frac{dr}{r})_\beta,
\ee
i.e. $\eta=\frac{1}{r^2}g_{2n+1}(\xi,\cdot)$.
We will restrict our considerations to no-where vanishing $e^B$.
Compatible with the cone metric \reef{cone1}, we can decompose the $SU(n+1)$ structure $J,\Omega$ via
\bea
J&=&r\eta\wedge dr+r^2e^{B}J_T\nn
\Omega&=&r^ne^{nB/2}(dr-ir\eta)\wedge\bar\Omega_T,
\eea
where $J_T$ is a two-form and $\bar\Omega_T$ is an $n$-form
both orthogonal to $\xi$ and $\partial_r$.
The conditions on the $SU(n+1)$ structure \reef{con} now become
\bea\label{combi}
dJ_T&=&0\nn
(J_T)^n\wedge d e^B&=&0\nn
-\frac{1}{c}e^B(J_T)^n+n(J_T)^{n-1}d\eta&=&0\nn
d\bar \Omega_T&=&\frac{i}{c}\eta\wedge\bar\Omega_T,
\eea
with the two-form $F$ given by
\be
F=-\frac{1}{c}J_T+d(e^{-B}\eta)
\ee
and we have introduced the constant $c=(n-2)/2$.
The Bianchi identity for $F$ is automatically satisfied and
so we just need to impose the equation of motion for $F$, \reef{Feom2},
to ensure that all equations of motion \reef{eom2}
are satisfied. From these conditions one can show that
\be
{\cal L}_\xi e^B=
{\cal L}_\xi J_T={\cal L}_\xi \eta=0,\qquad
{\cal L}_\xi \bar\Omega_T=\frac{i}{c}\bar\Omega_T.
\ee
From this we deduce that 
\be
{\cal L}_\xi J=0,\qquad {\cal L}_\xi \Omega=\frac{i}{c}\Omega
\ee
and hence that $\xi$ is Killing and holomorphic. One can also show
that $r\partial_r$ is holomorphic.

Let us now consider the link $L$, defined as $r=1$. The vector field $\xi$ restricts to a vector field
on $L$, which we denote by the same letter. Similarly, we find that $\eta$ and
$J$ pull back to well defined forms on $L$. One has to be a little careful about $\bar\Omega_T$:
it does not give rise to an $n$-form but rather a section of $\Lambda^n (T^*X)$ twisted by
the complex line bundle defined by $dr-ir\eta$. For this reason 
we have a $U(n)$ structure on $L$ (or a metric contact structure - see \cite{bg} definition 5).

We now introduce local coordinates on $L$ so that we can write $\xi=(1/c)\partial_z$, $\eta=c(dz+P)$ and
\bea\label{clang1}
ds^2_{2n+1}&=&c^2(dz+P)^2+e^{B}ds^2_{2n}.
\eea
Using \reef{combi} we deduce that $ds^2_{2n}$ is a local K\"ahler metric 
with K\"ahler-form $J_T$. Furthermore, defining $\Omega_T=e^{-iz}\bar\Omega_T$, 
we have that $\Omega_T$ is the local $(n,0)$-form on the K\"ahler manifold satisfying 
$d\Omega_T=iP\wedge \Omega_T$
and $dP=\rho$, where $\rho$ is the Ricci-form of the K\"ahler metric. We also write the
Ricci tensor of this K\"ahler metric as $R_{ij}$ and the Ricci scalar as $R$. 
The scalar field and the two-form then take the form
\bea\label{clang2}
e^{B}&=&c^2\left(\frac{R}{2}\right)\nn
F&=&-\frac{1}{c}J_T+cd[e^{-B}(dz+P)].
\eea
The equation of the two-form in \reef{Feom2} implies that the K\"ahler metric must satisfy
\be\label{keq}
\Box R+R_{ij}R^{ij}-\frac{1}{2}R^2=0.
\ee
At this point it is worth pausing to  emphasise that we have shown that
if we have a local K\"ahler metric in $2n$-dimensions that solves this master 
equation\footnote{Note that this equation first appeared, for $n=2$, 
in a different context in \cite{Cariglia:2004kk}.}
we can reconstruct a local $2n+1$-dimensional geometry via \reef{clang1}, \reef{clang2}
which admits solutions to the Killing spinor equations \reef{kspin2} and solves
the equations of motion \reef{eom2}.

Returning to the cone, in terms of this local description the original $SU(n+1)$ structure 
$J$, $\Omega$ is given by
\bea
J&=&-crdr\wedge(dz+P)+r^2e^{B}J_T\nn
\Omega&=&e^{iz}(e^{B/2}r)^n[dr-irc(dz+P)]\wedge
\Omega_T
\eea
and one can directly check that this $SU(n+1)$ structure satisfies \reef{con}.
One can also directly check that the equation of motion \reef{pink} is also satisfied: 
to do so observe that 
\be
e^{2(n-3)\phi}*_{2n+2}d[e^{2\phi}J]=
- c^2\left[\frac{(J_T)^{n-2}}{(n-2)!}\wedge \rho\wedge(dz+P)+*_{2n}d\left(\frac{R}{2}\right)\right].
\ee
Note that the natural orientation on the cone is given by
\be
\frac{(J)^{n+1}}{(n+1)!}= -cr^{2n+1}e^{nB}dr(dz+P)\frac{(J_T)^n}{n!}
\ee
and hence we take $\epsilon_{rzi_1...i_{2n}}=-1$ and we also take $\epsilon_{zi_1\dots i_{2n}}=+1$.

Since the Killing vector $\xi$ 
is no-where vanishing it defines a foliation. Just as in the case of Sasaki structures there
are three cases to consider. The regular case is when the orbits of $\xi$ close and the circle action is free.
In this case, the local description above is globally defined. In particular, it is characterised by K\"ahler
manifolds satisfying \reef{keq}. The quasi-regular case is when the
orbits of $\xi$ close but the action is only locally free. In this case the orbit space is an orbifold and 
$L$ is the total space of an orbifold circle bundle over a K\"ahler orbifold satisfying \reef{keq}. 
The irregular case is when the orbits generically do not close and there is no globally defined K\"ahler
geometry.

In the remaining sections we will illustrate the geometry we have introduced both in $2n+1$ dimensions
and $2n+2$ dimensions by discussing several ansatz, some of which lead to new explicit examples. 

\section{Fibration construction using $KE^+_{2n-2}$ spaces}

In this section we will construct explicit examples of the geometries
in $2n+1$ dimensions. For each K\"ahler-Einstein manifold with positive curvature in
$2n-2$ dimensions, the construction gives countably infinite classes of
simply connected, compact  geometries. 
The strategy is to first find
local K\"ahler metrics satisfying \reef{keq} and then afterwards show
that they lead to globally defined complete geometries in $2n+1$ dimensions.
The approach is the analogue of the construction of Sasaki-Einstein manifolds in
\cite{Gauntlett:2004yd,Gauntlett:2004hh} and generalises the analysis of 
\cite{gkw} from $n=3,4$ to all $n\ge 3$.

In order to find explicit examples of local K\"ahler metrics in $2n$ dimensions
satisfying~\reef{keq}, following \cite{Page:1985bq}, we consider the ansatz
\be
\label{ansatzlb}
ds^2_{2n} = \frac{d\rho^2}{U} + U \rho^2 (D\phi)^2
   + \rho^2 ds^2(KE^+_{2n-2})
\ee
with
\be
D\phi=d\phi+C.
\ee
Here $ds^2(KE^+_{2n-2})$ is a $2n-2$-dimensional K\"ahler-Einstein metric of
positive curvature. It is normalised so that ${\cal R}_{KE} = 2n
J_{KE}$ and the one-form form $C$ satisfies $dC=2J_{KE}$. Note
that $nC$ is then the connection on the canonical bundle of the
K\"ahler-Einstein space.
Let $\Omega_{KE}$ denote a local $(n-1,0)$-form, unique up rescaling by
a complex function.

To show that $ds^2_{2n}$ is a K\"ahler metric observe that the
K\"ahler form, defined by
\be
J_T = \rho d\rho \wedge D\phi + \rho^2 J_{KE},
\ee
is closed, and that the holomorphic $(n,0)$-form
\be
\Omega_T = e^{in\phi}
   \left(\frac{d\rho}{\sqrt{U}}+i\rho\sqrt{U}D\phi\right)
   \wedge \rho^{n-1} \Omega_{KE}
\ee
satisfies
\bea
d\Omega_T  = i f D\phi \wedge \Omega_T,
\label{rp}
\eea
with
\bea
f &=& n(1-U) - \frac{\rho}{2} \frac{dU}{d\rho}.
\eea
This implies, in particular, that the complex structure defined by
$\Omega$ is integrable. In addition~\eqref{rp} allows us to obtain the
Ricci tensor of $ds^2_{2n}$:
\be
{\cal R}=dP,\qquad P=fD\phi.
\ee
The Ricci-scalar is then obtained via $R={\cal R}_{ij}J^{ij}$.

We would like to find the conditions on $U$ such that $ds^2_{2n}$
satisfies the equation \reef{keq}. It is convenient to introduce
the new coordinate $x=1/\rho^2$ so that
\begin{equation}
\label{x-ans}
   ds^2_{2n} = \frac{1}{x}\left[
      \frac{dx^2}{4x^2U} + U (D\phi)^2 + ds^2(KE^+_{2n-2}) \right]
\end{equation}
and
\bea
\label{fR}
f &=& n(1-U)+ x \frac{dU}{dx}\\
R &=& 4(n-1) x f - 4x^2 \frac{df}{dx}.
\eea
We can now show that~\reef{keq} can be integrated once to give
\be
2(n-1) f^2 + U \frac{dR}{dx} = (constant)\times x^{n-2}.
\label{ans}
\ee

It is now straightforward to obtain polynomial solutions of
\reef{ans}. For simplicity we will restrict our considerations\footnote{For 
$n=3$ and $n=4$, see \cite{gkw} for some discussion of other solutions.} 
to solutions of the form $U=1-\alpha x^{n-2} (x-\beta)^2$.
Note that if we scale $x\to k x$, we obtain the same $2n+1$ dimensional
metric (see below) providing that $\alpha \to k^n\alpha$, $\beta\to \beta/k$. For reasons that will
become clear soon, we are interested
in $U$ having two distinct roots. If $n$ is even then we must have $\alpha>0$. If
$n$ is odd, by rescaling $x$ if necessary, we can also take $\alpha>0$.
We will also use this scaling to set $\beta =1$. Thus we will focus on solutions
with 
\be
U=1-\alpha x^{n-2} (x-1)^2.
\label{sol}
\ee
Observing that $U$ has turning points at $x=(n-2)/n$ and at $x=1$, 
we will choose $\alpha\in (\alpha_0,\infty)$ where
$\alpha_0=n^n/(4(n-2)^{n-2})$ so that $U$ has two positive roots 
$x_1$ and $x_2$.

We now consider the local metrics in $2n+1$ dimensions that can be constructed 
from these local $2n$-dimensional K\"ahler metrics:
\bea
ds^2_{2n+1}&=&c^2\left[(dz+P)^2+\frac{R}{2}ds^2_{2n}\right],
\eea
where $R$ is the Ricci-scalar of the $2n$-dimensional metric 
given in \reef{fR}:
\be
R=8\alpha x^{n-1}. 
\ee
The scalar $B$ and the two-form $F$ can be obtained from \reef{clang2}.
It will be very convenient to employ the coordinate transformation
\be
\phi=\frac{1}{n}(\psi-z)
\ee
so that the metric can be written
\be\label{metrew}
\frac{1}{c^2}ds^2_{2n+1}=wDz^2+\frac{RU}{2n^2wx}D\psi^2
+\frac{R}{8x^3U}dx^2+\frac{R}{2x}ds^2(KE^+_{2n-2})
\ee
where $D\psi=d\psi+nB$, $Dz=dz+g D\psi$ with
\bea
w&=&(1-\frac{f}{n})^2+\frac{RU}{2n^2x}\nn
g&=&\frac{1}{n^2w}(nf-f^2-\frac{RU}{2x}).
\eea
We demand that $w>0$, $R>0$ and $U\ge 0$. We will achieve this by demanding
that $x\in [x_1,x_2]$ where $x_i$ are two positive roots of $U(x)$. 

We now want to argue that this local metric, for countably infinite values of 
$\alpha\in (\alpha_0,\infty)$ and for
any positively curved K\"ahler-Einstein manifold,
globally extends to give a complete, compact metric on a $2n+1$ dimensional manifold. We will argue this in
two steps. We first study the $2n$-dimensional metric transverse to the $z$ direction in
\reef{metrew} and then consider the $U(1)$ fibration, with fibre parametrised by the
coordinate $z$, which we will take to be periodic with a suitable chosen period.

We start by analysing the $2n$-dimensional metric transverse to the $z$ direction in
\reef{metrew}. As we have already mentioned we take $x\in [x_1,x_2]$
and we take $\psi$ to be a periodic co-ordinate with
period $2\pi$.  Then the above $2n$-dimensional metric extends
to a smooth complete metric on the total space of an $S^2$ bundle over the original K\"ahler-Einstein space,
$KE^+_{2n-2}$, and this is true for any value of $\alpha\in (\alpha_0,\infty)$. 
To see this, we observe that the two sphere is parametrised by $\psi,x$. The key issue is
to ensure that the metric has no conical singularities at the poles of the two-sphere 
which are located at $x=x_1,x_2$. 
A small calculation show that since at any root $x_i$ of $U$ we have
\be
\frac{(U' x)^2}{w}|_{x=x_i}=n^2,
\ee
there will be no conical singularities if we take $\psi$ to have period $2\pi$.
Let us call this $2n$-dimensional manifold $Y_{2n}$.

We now turn to the $2n+1$-dimensional metric. The idea is
to choose $z$ to have period $2\pi l$, for suitable $l$, so that the metric is
that of the total space of a $U(1)$ fibration over the globally defined $2n$ dimensional base manifold
$Y_{2n}$, with connection one-form given by $l^{-1}gD\psi$. The key point here is to ensure that the periods of the
$(1/2\pi l)d(gD\psi)$ over a basis for the free part of the second homology group of $Y_{2n}$ are integers.
This is an almost identical set up to that discussed in \cite{Gauntlett:2004hh} and we refer to that paper for more details.
The result, in order to obtain a simply connected manifold, is that we need to choose
\be\label{cond}
\frac{g(x_2)}{g(x_1)}=\frac{p}{q},
\ee
where $p,q$ are relatively prime integers.
We also choose 
\be
l=\frac{hg(x_1)}{q},
\ee
where $h$ is the highest common factor of $p-q$ and $qc_i$ where
$c_i$ are the Chern numbers of the K\"ahler-Einstein manifold $KE_{2n-2}$.
Finally, we will show below that as $\alpha$ ranges from $\alpha_0$ to $\infty$,
$\frac{g(x_2)}{g(x_1)}$ monotonically increases from $0$ to $1$. Hence there
will be countably infinite values of $\alpha$ that satisfy \reef{cond}
and hence countably infinite metrics on complete $2n+1$-dimensional manifolds.

To examine the behaviour of $g(x_2)/g(x_1)$ as a function 
of $\alpha$ we first note that using 
$U(x_i)=0$ we obtain the simple expression
\be
g(x_i) = -\frac{2}{n(x_i-1)+2}.
\ee
To proceed, we recall that the turning points of $U(x)$ are at $x=(n-2)/n$ and
at $x=1$ and hence $(n-2)/n<x_1<1<x_2$. We thus conclude that $g(x_i)$ is
negative for both $x_1$ and $x_2$. We now observe that 
\be
\frac{d}{d\alpha} g(x_i) = \frac{2n}{(n(x_i-1)+2)^2} \frac{dx_i}{d\alpha}.
\ee
Next, since $U(x_i)=0$, we have 
\be
\alpha= \frac{1}{x_i^{n-2}(x_i-1)^2}
\ee
and we can compute
\be
\frac{dx_i}{d\alpha} = - \frac{x_i^{n-1}(x_i-1)^3}{n(x_i-1)+2}
\ee
This is negative for $x_2$, and positive for $x_1$ and hence we deduce the
required monotonicity property of $g(x_2)/g(x_1)$.

\section{An ansatz for the geometry in $2n+2$ dimensions}

Geometries in $2n+1$-dimensions that can be constructed from
a $2n$-dimensional K\"ahler manifold consisting of a product of K\"ahler-Einstein manifolds 
satisfying \reef{keq} were studied in \cite{gkw}. As we have discussed these give rise to 
conical geometries in $2n+2$-dimensions with an $SU(n+1)$ structure satisfying \reef{icon}, \reef{ipink}. 
In this section we will consider an ansatz for the geometry in $2n+2$-dimensions that generalises
these conical geometries. We will derive the ordinary differential equations that need to
be solved and while we have not been able to find the most general solution, we do 
recover some known solutions.

Our metric ansatz is given by
\be
ds^2_{2n+2} = \alpha^2 dr^2 + \beta^2 (dz+P)^2 + \sum_{i=1}^n \gamma_i^2
ds^2_i (KE_2),
\ee
where $ds^2_i(KE_2)$ denotes the metric of the $i$th two-dimensional K\"ahler-Einstein
space. We will let $J_i$, $\Omega_i$ and ${\cal R}_i$ denote the K\"ahler-form, the $(1,0)$ form 
and the Ricci-form of the corresponding K\"ahler-Einstein space, respectively. 
We also have $P=\sum P_i$ where
\be
dP_i={\cal R}_i = l_i J_i
\ee
and $l_i$ are constant.
Note that if we set some of $l_i$'s to the same value, and also set the corresponding
$\gamma_i$'s to be equal, one can also replace the relevant product of the two-dimensional 
K\"ahler-Einstein spaces with higher-dimensional K\"ahler-Einstein spaces.

The $SU(n+1)$ structure $J,\Omega$ is given by
\bea
J &=& - \alpha \beta dr \wedge (dz+P)
+ \sum_i \gamma^2_i J_i
\\
\Omega &=& \left[ \alpha dr - i \beta (dz+P)\right] \wedge \prod_i \gamma_i\Omega_i.
\eea
We will demand that the $SU(n+1)$ structure satisfies \reef{icon}, \reef{ipink} which
we write again here:
\bea
d(e^{n\phi}\Omega ) &=& 0
\\
d(e^{2(n-1)\phi} J^n ) &=& 0
\\
d\left[ e^{2(n-3)\phi} *_{2n+2} d(e^{2\phi} J ) \right] &=& 0.
\eea
%
If we set $e^\phi=\lambda$ and assume that $\alpha,\beta,\gamma_i$ and $\lambda$ are
all functions of $r$, we obtain the following set of coupled nonlinear ordinary
differential equations:
\bea
\alpha \gamma \lambda^n  + (\beta \gamma \lambda^n)' &=& 0
\\
\alpha\beta\gamma^2\lambda^{2(n-1)} \sum \frac{l_i}{\gamma^2_i}
+ ( \gamma^2 \lambda^{2(n-1)} )' &=& 0
\\
 \frac{\beta\gamma^2\lambda^{2(n-3)}}{\alpha\gamma^4_i}
 \left[ l_i \alpha\beta \lambda^2 + (\gamma^2_i\lambda^2)' \right]
 &=& k_i
 \\
 \sum k_i l_i &=& 0,
\eea
where $k_i$ are constants and $\gamma= \prod \gamma_i$.

To recover the simple metric cone geometries whose $2n$-dimensional base geometries were discussed in \cite{gkw},
we start with
\be
\alpha=1, \quad
\beta=\gamma_i= c r , \quad
\lambda=r^{k}.
\ee
The above equations then give
\be
c = \frac{n-2}{2}, \quad
k = -\frac{n-1}{n-2}, \quad
k_i = c^{2(n-1)}(l_i-1)
\ee
and
\be
\sum_i l_i = \sum_i l_i^2 = 1.
\ee
The base of this metric cone is the $2n+1$-dimensional geometry built from
a $2n$-dimensional K\"ahler base consisting of a product of K\"ahler-Einstein spaces.
The last equation, which up to a scaling was noted in \cite{gkw}, is the
algebraic form of the master equation \reef{keq} for the K\"ahler base consisting of
a product of K\"ahler-Einstein spaces.

Let us now construct some other solutions for $n=3$ and $n=4$ which give rise
to supersymmetric solutions of type IIB supergravity and $D=11$ supergravity via
\reef{10ant} and \reef{11ant}, respectively. We first consider the $n=3$ case.
The solution describing the $AdS_3$ limit of D3-branes wrapped on $H_2/\Gamma$, a Riemann surface with
genus $g>1$, in a Calabi-Yau four-fold \cite{Klemm:2000nj,Maldacena:2000mw,Naka:2002jz,Gauntlett:2006qw}, 
which was discussed from the present point of view in \cite{gkw}, can be recovered by setting 
\bea
l_1 &=& -1/3, \quad l_2 =l_3= 2/3,\nn
k_1 &=& -1/12, \quad k_2 =k_3= -1/48
\eea
and
\be\label{fgb}
\alpha=1, \quad \beta=\gamma_i=r/2, \quad \lambda=1/r^2.
\ee
Since $l_2=l_3$ and $\gamma_2=\gamma_3$ we can replace the 
corresponding $KE_2\times KE_2$ with a $KE_4$ space. 
For simplicity we will just discuss the case when $KE_4=CP^2$.

We would like to know whether there exists a more general solution to
the above coupled nonlinear differential equation, with the same
parameters $l_i,k_i$. One can indeed find that 
\bea
\alpha^2 &=& \left( 1 - \frac{m}{r^{4/3}} \right)^{-1}
\\
\beta^2 &=& \frac{r^2}{4} \left( 1 - \frac{m}{r^{4/3}} \right)
\\
\gamma_1^2 &=& \frac{r^2}{4}
\\
\gamma_2^2=\gamma_3^2 &=& \frac{r^2}{4} \left( 1 - \frac{m}{r^{4/3}} \right)
\\
\lambda^{-1} &=& {r^2 \left( 1 - \frac{m}{r^{4/3}} \right) },
\eea
satisfies the equations, where $m$ is a constant. Using \reef{10ant} 
we find that the ten dimensional type IIB metric takes the form
\bea
ds^2 &=& \frac{1}{r^2(1-\frac{m}{r^{4/3}})} ds^2(\bbR^{1,1} )
+
\frac{1}{4(1-\frac{m}{r^{4/3}})} ds^2(H_2/\Gamma) + \frac{dr^2}{r^2(1-\frac{m}{r^{4/3}})^2}
\nonumber
\\
&& + \frac{1}{4} \left[
(dz+P)^2 + ds^2(CP^2).
\right]
\eea
If we take $m>0$, $0\le r\le m^{3/4}$ we essentially have the solution of \cite{Klemm:2000nj,Maldacena:2000mw}.
Note that, as discussed in section 6.1 of \cite{Gauntlett:2006qw}, we can choose the range of $z$ to be $6\pi$ if $g-1$ is divisible
by three and $2\pi$ otherwise. In the former case, the solution interpolates between 
a locally $AdS_5\times S^5$ region and the $AdS_3\times H_2/\Gamma\times S^5$ solution given above in \reef{fgb}.
In the latter case the $S^5$ is replaced with $S^5/Z_3$.

The above solution can be interpreted as describing the near horizon limit of a $D3$-brane
wrapping a holomorphic $H_2/\Gamma$ inside a Calabi-Yau four-fold ($CY_4$) \cite{Maldacena:2000mw}. 
It would be interesting if we could find a solution that interpolated from an asymptotic $CY_4$ region to this solution
or perhaps just to the $AdS_3\times H_2/\Gamma$ solution \reef{fgb}. We have not been able to construct such a solution, but
we observe that our ansatz does include the following Calabi-Yau four-fold metric 
\be
ds^2=\frac{1}{U}dr^2+\frac{9}{16}Ur^2(dz+P_1+P_2)^2
+\frac{l_13}{8}(r^2+c)ds^2_1(KE_2)+\frac{9}{4}r^2ds_2^2(CP^2),
\ee
where 
\be
U=\frac{3r^2+4c}{9(r^2+c)},
\ee
we have normalised $ds^2(KE_2)$ so that $l_1=\pm 1$ and
we have taken $KE_4$ to be $CP^2$ with
$l_2=6$. When $l_1=1$, $KE_2$ is a unit radius $S^2$. Choosing
$c>0$ and $0\le r<\infty$ we have the metric of
\cite{Cvetic:2000db} (eq. (5.34)). In particular the range of $z$ is $2\pi$ and as $r\to \infty$
the metric is asymptotically a cone over a regular Sasaki-Einstein space, while
at $r=0$ we have an $S^2$ bolt: note that since the period of $z$ is $2\pi$ at
$r=0$ the $CP^2$ and the $z$ fibre combine to give $S^5/Z_3$ and so there is a conical singularity.

On the other hand when $l_1=-1$, the case of more relevance here, we choose $KE_2=H/\Gamma$, again a Riemann surface with genus $g>1$.
We also take $c<0$ and $0\le r^2<c$. We can choose the range of $z$ to be $6\pi$ if $g-1$ is divisible
by three and $2\pi$ otherwise. In the former case, there is no conical singularity at the $H^2/\Gamma$ 
bolt at $r=0$, while in the latter case there is. Note that this metric is singular as $r^2\to -c$. This metric
provides a natural local model of a holomorphic $H_2/\Gamma$ in a $CY_4$ for which $D3$-branes can wrap.

Now let us consider the M-theory case with $n=4$ which is very similar. 
Setting
\bea
l_1 &=& -1/2, \quad
l_2 =l_3=l_4= 1/2,\nn
k_1 &=& -3/2, \quad
k_2 =k_3=k_4= -1/2,
\eea
we obtain the solution
\be\label{fgb2}
\alpha=1, \quad
\beta = \gamma_i = r, \quad
\lambda = r^{-3/2},
\ee
which corresponds to the $AdS_2$ limit of M2-branes wrapping a holomorphic $H^2/\Gamma$, again a 
Riemann surface of genus $g>1$, in a Calabi-Yau five-fold \cite{Caldarelli:1998hg,Gauntlett:2001qs}.
We can also find a more general solution, with
\bea
\alpha^2 &=& \frac{1}{1-\frac{m}{r}}
\\
\beta^2 &=& \left( 1 - \frac{m}{r} \right) r^2
\\
\gamma_1^2 &=&  r^2
\\
\gamma_2^2=\gamma_3^2=\gamma_4^2 &=& \left( 1 - \frac{m}{r} \right) r^2
\\
\lambda^{-4/3} &=&  \left( 1 - \frac{m}{r} \right) r^2.
\eea
The corresponding $D=11$ metric can be easily constructed from \reef{11ant} and is given by
\bea
ds^2 &=& -\frac{1}{r^2(1-m/r)} dt^2 + \frac{1}{1-m/r} ds^2(H_2/\Gamma) +
\frac{dr^2}{r^2(1-m/r)^2}
\\
&& + (dz+P)^2 + ds^2(CP^3).
\eea
Where, for simplicity, we have restricted attention to the case of $KE_6=CP^3$. 
If we take $m>0$, $0\le r\le m$ we essentially have the solution of \cite{Caldarelli:1998hg,Gauntlett:2001qs}.
Note that we can choose the range of $z$ to be $8\pi$ if $g$ is odd
and $4\pi$ if $g$ is even. In the former case, the solution interpolates between 
a locally $AdS_4\times S^7$ region and the $AdS_2\times H_2/\Gamma\times S^7$ solution given above \reef{fgb2}.
In the latter case the $S^7$ is replaced with $S^7/Z_2$.

Again it would be interesting if we could find a solution that interpolated from an asymptotic $CY_5$ region to this solution
or perhaps just to the $AdS_2\times H_2/\Gamma$ solution \reef{fgb2}. We have not been able to construct such a solution, but
we observe that our ansatz does include the following Calabi-Yau five-fold metric 
\be
ds^2=\frac{80}{U}dr^2+\frac{U}{5}r^2(dz+P_1+P_2)^2
+l_1 2(r^2+c)ds^2_1(H_2/\Gamma)+16r^2ds_2^2(CP^3)
\ee
where 
\be
U=\frac{4r^2+5c}{(r^2+c)},
\ee
we have normalised $ds^2(KE_2)$ so that $l_1=\pm 1$ and
we have taken $KE_6$ to be $CP^3$ with
$l_2=8$. When $l_1=1$, $KE_2$ is a unit radius $S^2$. Choosing
$c>0$ and $0\le r<\infty$ we have the metric in the general class of
\cite{Cvetic:2000db}. The range of $z$ is $4\pi$ and as $r\to \infty$
the metric is asymptotically a cone over a regular Sasaki-Einstein space, while
at $r=0$ we have an $S^2$ bolt: note that since the period of $z$ is $4\pi$ at
$r=0$ the $CP^3$ and the $z$ fibre combine to give $S^7/Z_2$ and so there is a conical singularity.
On the other hand when $l_1=-1$, the case of more relevance here, we take $KE_2=H/\Gamma$,
$c<0$ and $0\le r^2<c$. We can now choose the range of $z$ to be $8\pi$ if $g$ is odd
and $4\pi$ if $g$ is even. In the former case, there is no conical singularity at the $H^2/\Gamma$ 
bolt at $r=0$, while in the latter case there is. Note that this metric is singular as $r^2\to -c$, but nevertheless
provides a good local model of a holomorphic $H_2/\Gamma$ embedded in a $CY_5$, for which membranes can wrap.

\section{LLM inspired ansatz}\label{llm}
In this section we consider an ansatz that is motivated by the
results of Lin, Lunin and Maldacena (LLM) \cite{Lin:2004nb}. 
It was shown in \cite{nak1} and \cite{nak2} how one can recast the results of
LLM in terms of a local K\"ahler geometry in $2n$ dimensions
satisfying \reef{inakeq}, for $n=3$ and $n=4$. Here we extend this by constructing
an ansatz for general $n$. 

We start with the following ansatz for the local $2n$-dimensional K\"ahler metric
\be
ds^2 = \frac{dy^2}{U} + y^2 U (D\psi)^2 + \frac{f}{U} (dx^2_1+dx^2_2)
+ y^2 ds^2(KE_{2n-4}),
\label{ansatz}
\ee
where $ds^2(KE_{2n-4})$ is a K\"ahler-Einstein metric (possibly local),
$D\psi = d\psi + \sigma +V$, $\sigma$ is a one-form on the
K\"ahler-Einstein space, $V$ is a one-form on the two-dimensional space
spanned by $x_i$, $i=1,2$ and $U,f,V$ all depend on three coordinates $y,x_i$.

We take the K\"ahler form to be given by
\be
J_T = y dy \wedge D\psi + \frac{f}{U} dx_1 \wedge dx_2 + y^2 J_{KE},
\ee
where $J_{KE}$ is the K\"ahler form on the K\"ahler-Einstein space.
Then, demanding that $dJ_T=0$ we obtain the following conditions
\bea
d\sigma &=& 2 J_{KE}
\\
d_2 V &=& \frac{1}{y} \partial_y
\left( \frac{f}{U} \right) dx_1 \wedge dx_2,
\label{ka}
\eea
where $d_2\equiv dx^i\wedge \partial_i$.
In order to see if the complex structure is indeed integrable we need to
compute the derivative of the $(n,0)$-form $\Omega_T$ given by 
\be
\Omega_T = e^{ik\psi}\left( \frac{dy}{\sqrt{U}} + i y \sqrt{U} D\psi \right)
\wedge
\sqrt{\frac{f}{U}} ( dx_1 + i dx_2 )
\wedge y^{n-2} \Omega_{KE},
\ee
where $\Omega_{KE}$ is the $(n-2,0)$-form on $KE_{2n-4}$ which
satisfies
\be
d\Omega_{KE} = i k \sigma \wedge \Omega_{KE}.
\ee
The constant $k$ determines the normalisation of the 
$KE_{2n-4}$ space. In particular, we have ${\cal R}_{KE}=2kJ_{KE}$,
where ${\cal R}_{KE}$ is the Ricci-form of the KE space.
We now find that 
\be
d\Omega = i P \wedge \Omega,
\ee
with the Ricci potential given by
\be
P = \frac{1}{2} *_2 d_2 \left( \ln f \right)
- \frac{y^{2-n}U}{\sqrt{f}} \partial_y \left( y^{n-1} \sqrt{f} \right)
D\psi
+ k (d\psi + \sigma ),
\ee
provided that we impose
\be
\partial_y V = \frac{1}{y} *_2 d_2 \left( \frac{1}{U} \right).
\label{int}
\ee
The compatibility of (\ref{ka}) and (\ref{int}) leads to the
following equation:
\be
\Delta \left( \frac{1}{U} \right)
+ y \partial_y \left[ \frac{1}{y} \partial_y \left( \frac{f}{U} \right)
\right] = 0,
\label{ma}
\ee
where $\Delta = \partial^2_1+\partial^2_2$.

Having obtained the Ricci potential, the next step would be to compute
the Ricci tensor and see how the master equation (\ref{inakeq}) can
be satisfied. To simplify things, we first introduce 
a function $D$ defined via
\be
\frac{1}{U} = \frac{y}{2} \partial_y D.
\ee
We can now readily integrate (\ref{int}) to get
\be
V_i = \frac{1}{2} \epsilon_{ij} \partial_j D , \quad i,j=1,2.
\ee
In terms of $D$, (\ref{ka}) is now expressed as
\be
\Delta D + \frac{1}{y} \partial_y \left( f y \partial_y D \right) = 0.
\label{toda}
\ee
Furthermore, based on hints from \cite{Lin:2004nb} in the $n=3,4$ case, 
we now make the assumption that there exists a relation
\be
f = y^{2p} e^{qD},
\ee
which makes the master equation (\ref{inakeq}) identically satisfied. Here
$p,q$ are constants that are to be fixed in terms of other parameters $n,k$ 
that we have already introduced.

Noting that now
\be
\frac{1}{2} *_2 d_2 \ln f = qV,
\ee
we can rewrite the Ricci potential $P$ as
\be
P = \left[ k-q - (n+p-1) U \right] D\psi + (q-k) V.
\ee
It is now straightforward to obtain the Ricci-form $dP$ and from that
the Ricci scalar which takes the simple form
\be
R = \frac{4}{y^2} \left[
(k-q)(n-2) - q (n+p-1) - (n+p-1)(n+p-2)U
\right].
\ee
After some computation one can now check 
that (\ref{inakeq}) is satisfied, if we demand that
\be
p= 3-n
\ee
and
\be
(n-2)(q-k)[q(n-1)-k(n-3)]=0.
\ee

Let us first discuss the special case when $n=3$. In this case we
can solve the equations by taking $p=q=0$. Then the only equation
that needs to be solved is the linear equation
\be
\Delta D + \frac{1}{y} \partial_y ( y \partial_y D ) = 0
\ee
and we have recovered the equation of LLM for the 
type IIB case\footnote{In order to recover the LLM result one, when 
one adds in the extra coordinate $z$ to obtain a seven manifold. One
should also shift the $\psi$ coordinate $\psi\to \psi+\alpha z$ for 
some constant $\alpha$.}

For generic values of $n\ge 3$, the second equation is
satisfied if $q=k$. Since $k$ is related to the scalar curvature of
the K\"ahler-Einstein base we can rescale it to $0, \pm 1$ without
losing generality. Here let us assume that $k\neq 0$. Then the
entire solution is governed by the equation 
(when $k=-1$ we redefine $D\rightarrow -D$)
\be
\Delta D + \frac{1}{y} \partial_y ( y^{7-2n} \partial_y e^D ) = 0.
\label{gToda}
\ee
After the coordinate change $x=(2n-6)^{2(3-n)/(n-2)}y^{2n-6}$, this equation
becomes
\be
\Delta D + x^{\frac{n-4}{n-3}}\partial_x^2 e^D  = 0.
\ee
When $n=4$ this equation is the continuous Toda equation just as 
LLM discovered in the context of $D=11$ supergravity \cite{Lin:2004nb}.
We do not know whether or not (\ref{gToda}) is also an integrable
system for $n>4$. But it is at least clear that (\ref{gToda}) still enjoys
the 2d conformal symmetry, i.e. the equation is invariant under transformation
\be
x_1+i x_2 \rightarrow g(x_1+i x_2) , \quad\quad
D \rightarrow D - \log |\partial g |^2
\ee
for an analytic function $g$.

\section{Conclusions}

In this paper we have introduced a new class of geometries in $2n+2$ dimensions
that are specified by a metric, a scalar and a three-form. The geometries
admit a specific kind of Killing spinor or, equivalently, a specific kind
of $SU(n+1)$ structure. 
For $n=3$ and $n=4$ these give rise to
supersymmetric solutions of type IIB and $D=11$ supergravity, with
$\bbR^{1,1}$ and $\bbR$ factors, respectively.

We also showed that if these geometries in $2n+2$ dimensions are a 
certain kind of metric cone,
then we obtain a new class of metric contact 
geometries in $2n+1$ dimensions, on the base of the cone,
that are specified by a metric, a scalar and a two-form.
For $n=3$ and $n=4$ these give rise to
supersymmetric solutions of type IIB and $D=11$ supergravity, with
$AdS_3$ and $AdS_2$ factors, that were discussed in \cite{nak1}
and \cite{nak2}, respectively. We have noted the strong similarities with Ricci-flat
K\"ahler cones and Sasaki-Einstein manifolds.

We also constructed some specific examples of these geometries in sections 4-6. 
The constructions in section 4 can be straightforwardly extended
by generalising the construction of section 5 of \cite{gkw}. It should
also be possible to extend this construction further using the results of
\cite{Chong:2004ce,Chen:2007du}. In sections 5 and 6 of this paper our constructions
boiled down to solving some differential equations and we think it would
be worthwhile to try and find additional solutions.

\subsection*{Acknowledgements}
We would like to thank Jaume Gomis, Jan Gutowski, Oisin Mac Conamhna 
and Daniel Waldram
for helpful discussions. We thank the Perimeter Institute for hospitality
where this work was completed. NK would also
like to thank the Institute for Mathematical Sciences at
Imperial College for hospitality. 
NK is supported by the
Science Research Center Program of the Korea Science and Engineering Foundation (KOSEF) through the Center for Quantum Spacetime (CQUeST) of Sogang
University with grant number R11-2005-021. 
JPG is supported by an EPSRC Senior Fellowship and a Royal Society Wolfson Award.

\end{document}